# Plasmonic dielectric antennas for hybrid optical nanotweezing and optothermoelectric manipulation of single nanosized extracellular vesicles

*Chuchuan Hong1,2, Ikjun Hong1,2, Yuxi Jiang3,4, Justus C. Ndukaife\*1,2,5*

[1]Department of Electrical and Computer Engineering, Vanderbilt University, Nashville, TN, USA

[2]Vanderbilt Institution of Nanoscale Science and Engineering, Vanderbilt University, Nashville, TN, USA

[3]Department of Electrical and Computer Engineering, University of Maryland College Park, MD, USA

[4]Insitute for Research in Electronics and Applied Physics (IREAP), University of Maryland College Park, MD, USA

[5]Department of Mechanical Engineering, Vanderbilt University, Nashville, TN, USA

ABSTRACT: We present an experimental demonstration of near-field optical trapping and dynamic manipulation of a single extracellular vesicle using a plasmonic dielectric nanoantenna that supports an optical anapole state. The optical anapole is a non-radiating optical state generated by the destructive interference between electric and toroidal dipoles in the far-field. To enhance the trapping capabilities, we employ a plasmonic mirror to enhance the anapole state. By harnessing the enhanced electromagnetic hotspot resulting from the mirror-enhanced anapole state, we achieve a high trapping potential of approximately 3.5 $K_bT$. The dynamic manipulation of the vesicle is achieved by inducing a thermoelectric field in the presence of an ionic surfactant and the resulting plasmonic heating. Specifically, we introduce cetyltrimethylammonium chloride (CTAC) as the ionic surfactant and utilize the local heating generated by the plasmonic reflector to create a thermoelectric field. This enables active transport, stable trapping, and dynamic manipulation of a single extracellular vesicle. Moreover, the thermoelectric field contributes to an increase in the overall trapping potential.

The ability to trap and dynamically manipulate nanoscale particles and biomolecules in solution is essential in nanoscience and biology. Extracellular vesicles (EVs) are membranous particles released by cells containing important biological molecules such as proteins, nucleic acids, and lipids[1,2]. They are classified into two categories according to their biophysical properties and biogenesis mechanisms. They include exosomes and microvesicles[3]. Exosomes are 30 nm to 130 nm in diameter and derived from endosomal compartments. Microvesicles, on the other hand, are derived from the outward budding and pinching-off of the cell plasma membrane. Given the heterogeneity of EVs, there is a need to trap and dynamically manipulate individual EVs in solution. However, this need has yet to be met using the conventional optical tweezer technology. Trapping nanoscale EVs with optical tweezers requires substantial optical power that predisposes them to damage. For example, earlier works on laser trapping of EVs require at least 100 mW of input laser power and have reported the explosion of EVs while held in the trap[4]. Furthermore, optical tweezers do not guarantee single EV trapping, as they typically capture multiple EVs inside the traps due to the diffraction-limited width of the trapping potential well, which is considerably larger than the size of single nanoscale EVs.

Nanotweezers leveraging near-field enhancement and confinement present an alternative approach for trapping nanoscale objects with a few milli-Watts of laser power. The optical nanotweezers comprise plasmonic cavities, dielectric nanoantenna supporting Mie resonance, or photonic crystal cavities. Dielectric near-field tweezers based on silicon pillars[5], silicon metasurface[6], photonic crystal[7,8], or waveguides[9] have also been investigated. Several plasmonic nanotweezers using double nano-hole aperture[10–12], coaxial aperture[13,14], gold nanopillars[15,16], nano-pyramids[17], or nano-blocks[18] have also been reported for trapping nanoscale objects like protein molecules, nanoparticles, or DNA molecules.

However, there are several challenges with existing optical nanotweezers. The dielectric antennas and photonic crystal cavities do not offer dynamic manipulation capabilities. Traditional plasmonic nanotweezers suffer from local heating due to intrinsic plasmonic material loss, whereby excessive heating can be detrimental to delicate biological particles[19]. In addition, plasmon-induced heating often induces thermal-related hydrodynamic effects that interfere with the trapping stability, such as positive thermophoresis or buoyancy-driven convection[20–22]. In plasmonic nanoantennas, the optical absorption of an object scales with the intensity of the enhanced local[23,24], and the nature of plasmonic modes ensures that most of the light energy resides in the very vicinity of the plasmonic resonator. Thus, a plasmonic structure, which enhances the local light intensity and, in turn, the gradient trapping force, also serves as a strong nanoscale heat source due to the inherent material loss. Efforts to eliminate the temperature increase arising from the local heating effect have been investigated and reported using high thermal conductivity films or substrates that serve as heat sinks[25,26], as well as by employing off-resonant excitation[27,28].

On the other hand, local heating is not always inhibitive to trapping but can be turned to an advantage with the help of an applied a.c. electric field[16,29,30] or surfactants[31,32], relying on electrothermoplasmonic flow or thermoelectric effect, respectively. They both allow rapid particle transport towards the hotspot, as well as dynamic manipulation[22,31]. Thus, an elegant way to design near-field nanotweezers is to properly manage the local heating, so that local field enhancement can be as large as possible to exert strong optical force, while ensuring at the same time that the temperature rise remains moderate, to preclude any harm to trapped specimens but facilitating transport. A dielectric-plasmonic hybrid system should meet this requirement, wherein the peak electric field is defined by the dielectric nanoantenna and spatially separated from the plasmonic materials. The multipolar nature of dielectric modes also offers flexibility in design. Among them, the optical anapole has attracted our attention because of its potential to generate substantial field enhancement with a single antenna[33].

RESULTS AND DISCUSSIONS

Optical anapoles result from the far-field destructive interference of electric dipoles and toroidal dipoles. Thus, the light energy is highly confined in the near-field. Its substantial near-field enhancement has enabled several on-chip applications, including lasing[34], nonlinear optics[35], and strong coupling phenomenon[36]. Here, we propose and experimentally demonstrate our hybrid anapole-plasmonic tweezer (HAT), which uses an optimized optical anapole state to hold single EVs or 20 nm polystyrene beads. Most of the electric field enhancement is confined inside the anapole disk. The immediate solution to make the near-field accessible is to open a slot at the very center of the anapole nanoantenna[33,37]. The narrow slot, in turn, boosts the electric field due to the continuity of the normal component of the displacement field boundary condition. Moreover, due to the small refractive index of EVs (1.37-1.42)[38,39], low-power trapping of EVs requires a strong near-field enhancement. We thus further optimized the optical anapole field distribution by reshaping the slot into a double nano-hole (DNH) shape[40] and placing the silicon nanoantenna on a plasmonic-mirror engineered substrate. The plasmonic mirror comprises a 200 nm thick gold film with a 15 nm alumina spacer layer deposited on top of a glass chip. The use of the mirror (including gold film and spacer layer) largely boosts the near-field enhancement inside the DNH slot. The comparison between an anapole disk with a rectangle slot and the optimized anapole is provided in Fig. S1. This improved near-field enhancement stems from the fact that the mirror tunes the phase of the reflected light. Therefore, the reflected light and the incident light constructively interfere and create the maxima of the interfered wave amplitude at 130 nm above the spacer layer[41–43], which is near the top of the anapole disk. The precise positioning of the maximal electric field distribution not only boosts the field enhancement, but also optimizes the accessibility of the field to objects within the solution to be trapped, as shown in Fig. S2 d to f. A more detailed analysis of the purpose of adding the mirror and creating constructive interference is included in Fig. S2.

Part of the light in the near-field underneath the anapole disk is absorbed by the gold mirror, which induces a slight local temperature increase that can be utilized to actuate particle transport. It is evidenced by showing the simulated absorption cross-section peaks at the same wavelength as field enhancement reaches its maximum (shown in Fig. S3a and c). The thickness of the spacer layer needs to be carefully designed. A too-thin spacer layer (< 10 nm, for instance) generates a strong electric field inside the spacer layer between the bottom of the silicon disk and the gold reflector. This electric field is not accessible to the EVs but incurs significant absorption due to the Ohmic loss originating from the gold. Therefore, the field enhancement within the slot diminishes, as depicted in Fig. S3a and b. To avoid the reduction of optical force and overheating the trapping particle, the spacer layer's ultimate thickness is determined to be 15 nm. More details on the effects of the spacer layer thickness can be found in the SI section 1, and Fig. S2 and 3.

The continuous gold film serves not only as a heat source but also functions as a heat sink to dissipate excessive heat. The dual functionality enables improved management of the local temperature rise[44,45]. Under the same level of laser power, the temperature rise generated from the HAT system is commensurable with the reported values from previous plasmonic cavities[45–47]. However, the maximum intensity enhancement is at least 7 times stronger, indicating HAT system is more robust against the detrimental impact of local heating, such as positive thermophoresis[27,48].

By adding cetyltrimethylammonium chloride (CTAC), we exploit the thermoelectric effect[31,32] and turn the local heating into a contributor to trapping. The thermoelectric field enables rapid particle transport toward the hotspot and dynamic manipulation when we shift the laser from one anapole nanoantenna to another. The schematics of particle loading, trapping, and dynamic manipulation are illustrated in Fig. 1a.

Fig. 1b and c show the simulated near-field profiles on x-y plane and x-z plane, respectively. The incident light is polarized along the x direction and is propagating from the superstrate into the substrate. The simulation results suggest a highly confined near-field at the very center of the DNH slot, with the maximum field enhancement as high as 38. The optimized anapole state increases the field enhancement by 3.6-fold relative to the unoptimized anapole state (Fig. S1a and b), meaning more than an order of magnitude higher optical intensity and optical force can be generated from the optimized anapole state. Given this strong field enhancement, we numerically calculated the optical gradient force using the Maxwell's Stress Tensor (MST) formalism in a Finite-Difference Time-Domain solver (Ansys Lumerical). A 50 nm EV nanoparticle was assumed in the simulation. The refractive index of the EV was taken as 1.4, which is within the range of previously reported values[38,39]. The calculated trapping potentials along the x and z directions imply that stable trapping of EVs is expected from the optimized anapole state.

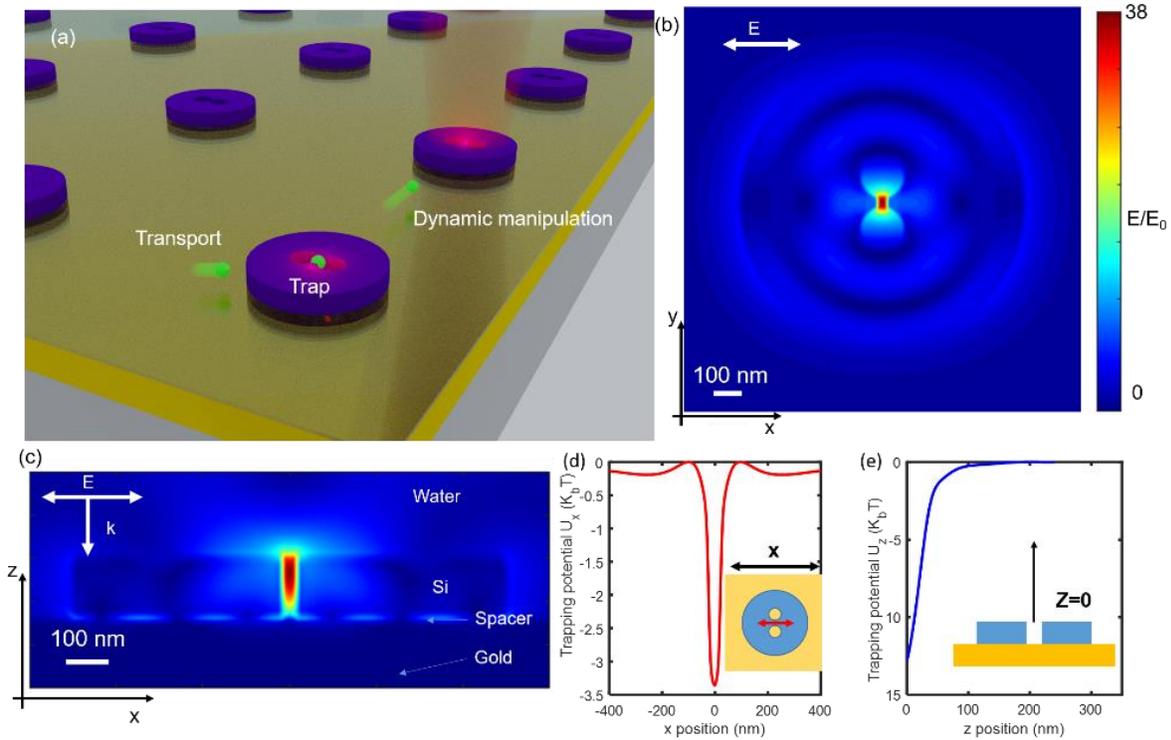

Figure 1: (a) is the schematic illustration of HAT system being able to trap, transport and dynamically manipulate single EV. (b) is the simulated in-plane electric field enhancement at the middle of silicon disk, where the polarization is oriented along the x direction. (c) is the simulated out-of-plane field enhancement cross the very center of the disk. Light is propagating along the opposite-z direction. (d) and (e) numerically calculated optical trapping potential given the electric field distributed at (b) and (c) under 7.2 mW/μm$^2$ laser illumination, respectively. The particle is assumed to be a 50 nm diameter EV (n=1.4).

The SEM diagram of a fabricated anapole is shown in Fig. 2a, where the tip-to-tip gap width is ~30 nm, matching the designed value in simulations in Fig. 1b and c. To maintain the small feature size, we adapted a double-layer photoresist electron-beam lithography process during the nanofabrication. More details about the fabrication are included in the supporting information. We then measured the scattering cross-section of a fabricated single silicon disk immersed in water using our homemade dark-field measurement setup. The measured data was normalized to its maximum value and plotted in Fig. 2b, labelled as the blue curve. Together we plot out the simulated scattering cross-section normalized to its own maximum in the orange curve. The measured data and the simulated spectrum match well, implying a strong field enhancement from our fabricated anapole nanoantenna. We notice that the scattering cross-section of this hybrid anapole system shows a Fano-like lineshape. Inside the spacer layer, in Fig. 1c, a clear Fabry Perot (FP) mode exists between the gold reflector and the bottom of the silicon disk. This FP mode couples to the original anapole state supported by the silicon disk and modifies the lineshape of the anapole spectrum. As depicted in Fig. 2b in red, the field enhancement at the very center of the DNH slot peaks at 973 nm, which is the wavelength of our trapping laser.

The initial trapping experiment was conducted using 10$^6$ EV/ml FITC-conjugated EVs (Creative Diagnostics) in pure DI water to demonstrate the ability of our HAT system to achieve near-field trapping. A 973 nm laser with a 10.6 mW/μm² intensity was focused on the anapole disk. An EV diffusing towards the anapole antenna was trapped, as shown in Fig. 2c. It took us, on average, ~40 minutes to observe an EV diffusing into the vicinity of the anapole nanoantenna and being trapped. The frame sequence in Fig. 2c shows the process of a single EV diffusing into the trap, staying trapped for over 90 seconds (the duration of the laser was on), and finally being released from the trap by turning off the laser. The whole process was recorded in supplementary video 1. The scatter plot of the particle's displacements extracted from the video is presented in Fig. 2d, which clearly shows improved confinement of the trapped EV under higher laser power, indicating the occurrence of near-field trapping. It should be noted that under this level of laser intensity of 10.6 mW/μm², single EV trapping using conventional laser tweezers has not been achieved, nor did we observe that in the experiment.

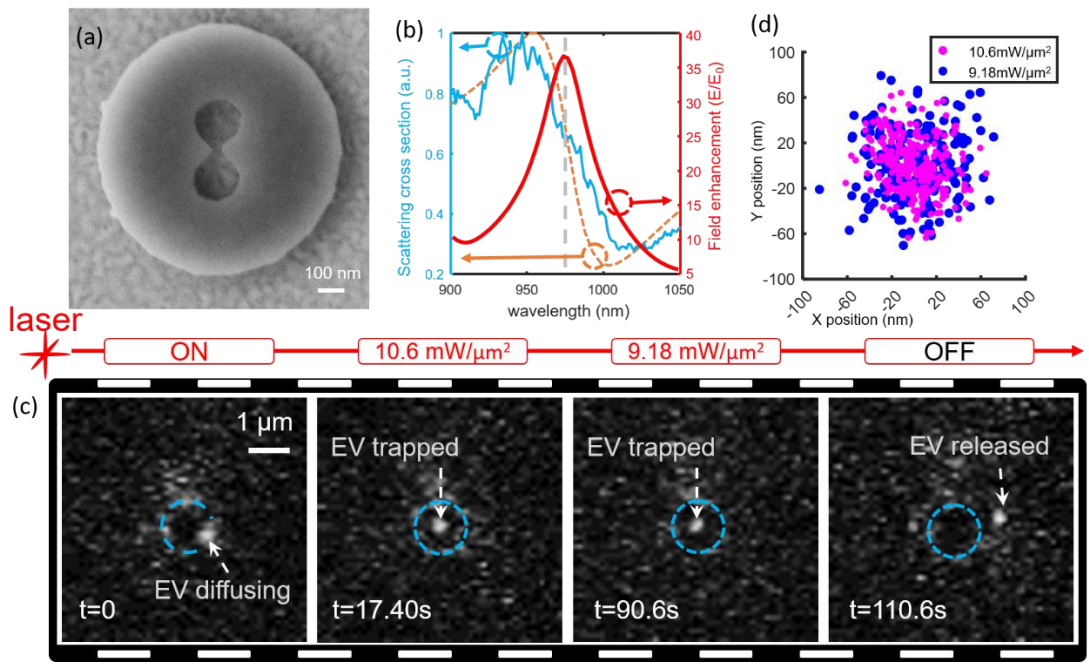

Figure 2: (a) is the SEM image of the fabricated anapole nanoantenna with ~30 nm tip-to-tip distance of the DNH slot. (b) depicts the simulated scattering cross section (orange dashed line) normalized to its maximum value and simulated field enhancement (red curve) monitored from the very center of the DNH slot. The blue curve represents the measured scattering cross section (normalized to its maximum value) from a homemade dark field setup. The simulation result matches well with the experimental measurements. The grey dashed line marks 973 nm trapping laser wavelength. (c) is the sequence of frames showing the diffusing, trapping, and releasing of a single EV suspended in DI water on the anapole nanoantenna. The blue dotted circle depicts the periphery of the silicon disk under the microscope, and the bright spot pointed by the white arrow is the fluorescence-labelled EV. The laser was focused onto the anapole nanoantenna for the first three frames through a water-immersed 60× objective lens with N.A.=1.2. (d) shows the scatter plot of the trapped EV positions under various laser intensities. As expected, higher laser power confines the trapped EV within a smaller area.

Having demonstrated near-field optical trapping using the HAT system, we proceeded to demonstrate dynamic manipulation and trapping by inducing and harnessing the thermoelectric field in the HAT system. The temperature rise in the HAT system is depicted in Fig. 3a. It is evident that the peak temperature rise is approximately 10 K, which corresponds to an absolute temperature of 35 °C at the center of the trap. We note that the human body temperature is 37 °C, which is the physiological temperature for extracellular vesicles that exist in human blood and plasma. Thus, this temperature rise in the HAT system is safe for biological objects.

The thermoelectric field is induced by adding CTAC. The final concentration of CTAC used is 1 mM. As the concentration of CTAC is beyond its critical micelle concentration (~0.13 mM), the positively charged $CTA^+$ ions agglomerate to form micelles. The micelles are thus also positively charged, leaving net negatively charged $Cl^-$ ions in solution. In the presence of local temperature gradients, both the negative $Cl^-$ and positive micelle ions experience positive thermophoresis and migrate away from the thermal hot spot. The micelles turn out to drift much faster than Cl$^-$[31]. Consequently, at steady-state, there exists a net spatial separation between micelles (positive) and Cl$^-$ (negative), resulting in a local thermoelectric field pointing towards the hot spot. EVs are known to have a negative surface charge[49,50]. The positively charged CTA$^+$ establishes a positive charge on the EV surface. Accordingly, the EVs move along the built-up thermoelectric field toward the hot spot. The mechanism of how the thermoelectric field is established and the EV migration is schematically depicted in Fig. 3a, where the simulated temperature profile under 7.2 mW/μm$^2$ laser intensity shows a 10 K maximum temperature rise.

We added 1 mM CTAC and repeated our experiment, placing the laser on the anapole nanoantenna. This time, we observed rapid particle transport to load the trap by taking advantage of the thermoelectric effect. In Fig. 3b, we illustrate how the experiment progressed. Initially, the EV was only making random Brownian motion in the vicinity of the anapole nanoantenna with no laser illumination. However, as the laser was focused on anapole 1, which is circled in blue, the EV started to move toward it. The transport finished after covering approximately 10 μm in 6.01 seconds, successfully capturing the EV by the anapole nanoantenna. At this point, the HAT system offers the following three options for the next step: (1) turning off the laser to release the EV; (2) keeping the laser on to hold the EV; or (3) dynamically manipulating the trapped EV by relocating the laser spot to a neighbor anapole nanoantenna, specifically anapole 2 in Fig. 3.

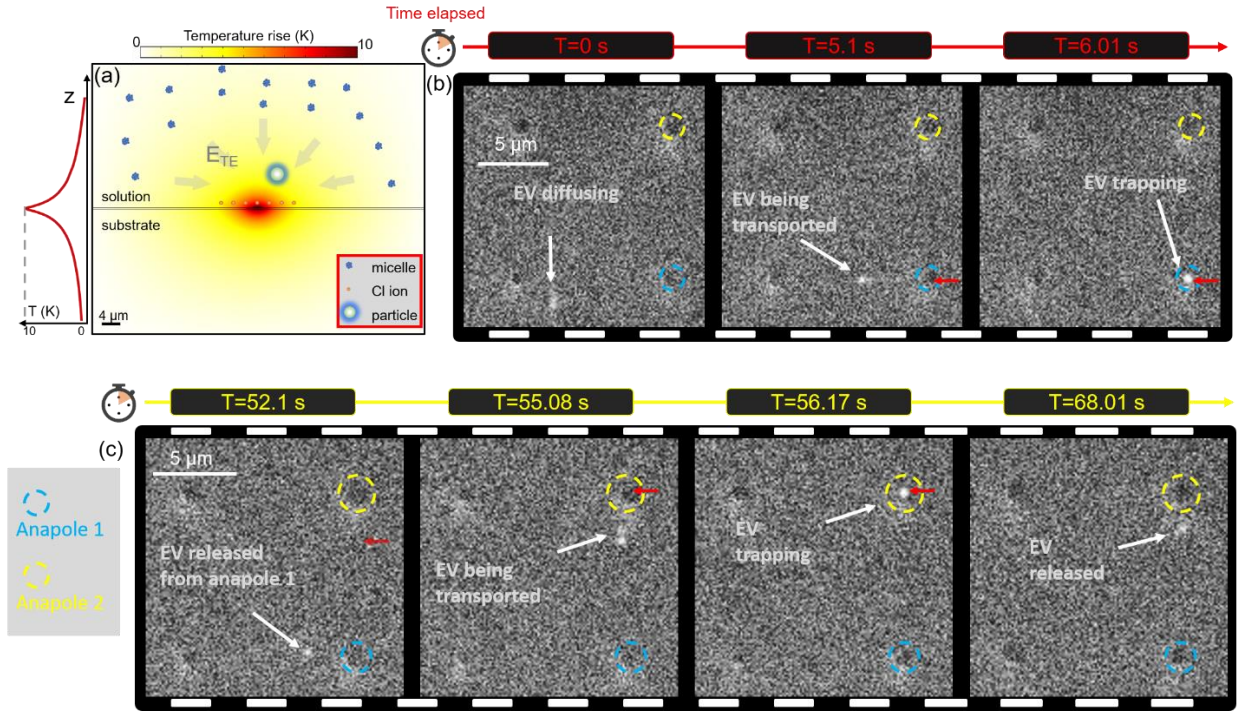

Figure 3: (a) simulated temperature profile using 7.2 mW/μm² laser illumination with 1.33 μm laser spot. $E_{TE}$ represents the local thermoelectric field, which drives the positively coated particle toward the hot spot. (b) and (c) sequences of frames elaborating the fast loading and dynamic manipulation process. The blue and yellow dashed circles highlight the positions of anapole 1 and 2, respectively. The red arrow indicates the position of laser illumination. The bright spot pointed to by the white arrow is the fluorescence-labelled EV.

The third option demonstrates another key capability of the HAT system, which is dynamic manipulation made possible by the induced thermoelectric field. To achieve this, we precisely shifted the laser spot from anapole 1 to anapole 2. As shown in the first frame of Fig. 3c, when the laser was not on an anapole nanoantenna, the EV was freely moving around in random Brownian motion. In the absence of an anapole state with the same laser illuminated on just gold film, the simulated temperature profile shows a mere 1.2 K temperature rise, as presented in Fig.

S6b. This temperature rise is an order of magnitude smaller than the temperature rise in the presence of the anapole state, making the thermoelectric effect negligible. However, as soon as the laser was aligned with anapole 2, the thermoelectric field is induced, and the EV started migrating, ultimately being captured by anapole 2 after 5.8 seconds. This whole process is presented in supplementary video 2.

We also observed that CTAC not only provides an approach for dynamic manipulation but also improves trapping stability. To better understand the influence of CTAC on the trapping stability, we performed the same transport and trapping experiments on fluorescence-labeled polystyrene beads with a diameter of approximately 20 nm (Thermo Fisher Scientific). The recorded videos are provided in supplementary video 3-5. This was done to avoid using heterogeneous EVs, which could introduce bias due to size differences. Under the same laser power on the same anapole nanoantenna, Fig. 4a clearly shows that the 20 nm polystyrene bead is confined within a smaller area in the presence of CTAC in solution. Fig. 4b and c are histograms of the particle position along the x and y directions, respectively, corresponding to Fig. 4a. They confirm our observation in Fig. 4a. Fig. 4d illustrates the simulated thermoelectric potential superimposed by the optical potential for a 20 nm polystyrene bead under 10 mW laser illumination on an anapole nanoantenna. The thermoelectric potential reaches approximately 4 $k_bT$ and covers a range of -/+ 20 μm, dropping to approximately 1 $k_bT$ nearly 10 μm away from the center. This matches the observed distance in the experiments where the particles started to perform an obvious directional movement. We also attribute the enhanced trapping stiffness to synergistic contributions from this thermoelectric potential. However, the main contribution to the trapping still comes from the nanoscale optical hotspots due to the anapole state, as shown in Fig. 4e. The optical potential makes the major contribution to the total potential by confining the particle within -/+ 40 nm, while the thermoelectric potential depth within this small range is shallow and not spatially confined. Although several other thermal-related effects have been reported in prior literature, such as buoyancy-driven convection[20], dispersion force[51], or thermo-osmosis[52], their contributions are trivial based on our numerical simulations. More details can be found in the SI sections.

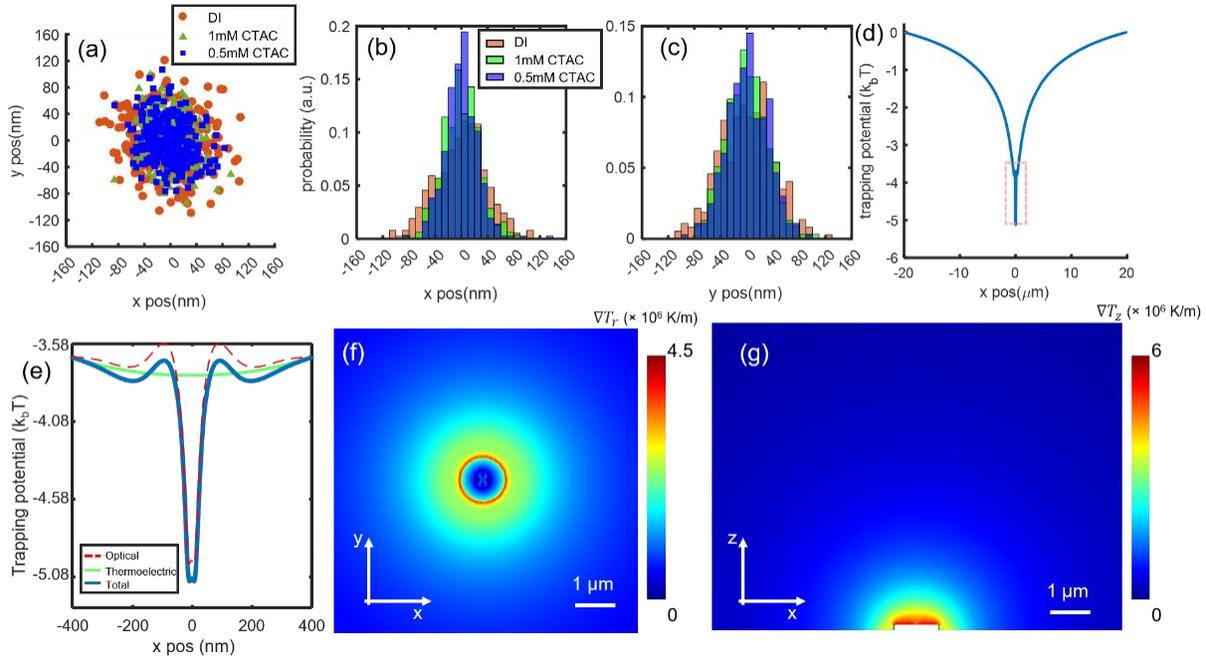

Figure 4. (a) is the scatter plot showing the positions of 20 nm polystyrene beads trapped in various solvent conditions. The stability inside CTAC clearly shows an improvement. (b) and (c) are the position histograms along x and y directions corresponding to (a). (d) shows the simulated superimposed trapping potential combining thermoelectric with optical potential in 1 mM CTAC concentration under 10 mW laser (7.2 mW/µm$^2$) illumination on a 20 nm polystyrene bead. (e) is a zoomed-in version of the thermoelectric potential, optical trapping potential, and total potential (optical + thermoelectric) within the region highlighted by the pink dotted box in (d). The optical potential is shown by the red dashed line. The thermoelectric potential is the solid green line, and total potential is the blue line. (f) is the simulated radial

temperature gradient on xy plane, 10 nm above the anapole top surface. (g) is the simulated vertical temperature gradient on xz plane across the middle of anapole silicon disk. (f) and (g) are both corresponding to the temperature distribution in Fig. 3a. The temperature gradient enables particles to be transported toward the anapole nanoantenna.

CONCLUSIONS:

To summarize, we demonstrated near-field trapping and dynamic manipulation enabled using an optical anapole state made possible by the substantial enhancement of the near-field. A series of optimizations were implemented to further amplify the field enhancement, such as the integration of a plasmonic mirror and the introduction of a double nanohole slot. As a result, we were able to successfully achieve the near-field trapping of single extracellular vesicles. Furthermore, the introduction of CTAC into the solution led to the establishment of a thermoelectric field under laser illumination, enabling rapid particle transportation and dynamic manipulation when the laser spot was shifted between adjacent nanoantennas. This HAT system provides a promising platform for a variety of compelling applications, including but not limited to single extracellular vesicle analysis[53,54], nano-plastics examination[55,56], or enhanced single-particle imaging and spectroscopy[33,57].

METHOD:

**Sample preparation:**

Lyophilized fluorescence labelled EVs are purchased from Creative Diagnostics. 100 μl DI water was added to 100 μg lyophilized EV to make a 1 μg/μL original solution. The sample was then diluted by DI water into the desired concentration for the trapping experiment in DI water.

100 μL diluted EV solution was then taken out and mixed with CTAC solution (Sigma Aldrich) to finally yield EV solutions with 1mM CTAC.

The fluorescence labelled polystyrene beads are purchased from Thermofisher. 200 μL $10^8$ PS/ml polystyrene bead solution was added with 200 μL 2 mM and 1 mM CTAC solution to make polystyrene beads solution with 1 mM and 0.5 mM CTAC, respectively, for the polystyrene beads trapping experiments.

To make the microfluid channel for trapping demonstration, after obtaining the fabricated anapole chip, we used double-sided spacer tape to create a 120 μm micro channel aroundthe chip, where solutions are injected in between the chip and cover slip.

**Fluorescence image:**

The fluorescence imaging and trapping experiments were performed using a custom fluorescent imaging and optical trapping microscope setup on a Nikon Ti2-E inverted microscope. The suspended particle solution was injected into the microfluidic channel mentioned in the last paragraph. A high quantum efficiency sCMOS (Photometrics PRIME 95B) camera was used to acquire the images. The anapole nanoantennas were excited by a 973 nm wavelength diode laser (Thorlabs CLD1015). The laser beam was focused with a Nikon 60X water-immersed objective lens (N.A.=1.2).

ASSOCIATED CONTENT

**Supporting Information**.

Supporting information: additional discussion on the role of plasmonic reflector, fabrication flows, single antenna dark-field measurement, thermal simulation, convection flow, method to calculate thermoelectric effect, dispersion force and thermos-osmosis simulation. (PDF)

**Supporting video 1: Trapping of single extracellular vesicle (EV) in DI water.**

Starting from a single EV undergoing random Brownian motion in water, we observed the EV diffuses into the trap and is captured. After that, we reduced the laser power and observed a decreasing trapping stability. We then turned off the laser and the EV was released from the trap. (MOVIE)

**Supporting video 2: Transport, trapping and dynamic manipulation of single EV in 1 mM CTAC solution.**

Starting from 'laser off' and a single EV making Brownian motion near the surface, we then turned on the laser on the anapole disk to establish thermoelectric field. As soon as the laser was on, the EV started a directional movement towards the laser illumination. It took about 6.01 seconds to cover ~10 μm distance, after which the EV was finally loaded onto the anapole nanoantenna. To perform dynamic manipulation, we relocate the laser spot onto a neighbor nanoantenna. During the translation of laser spot, the EV was temporarily released. Immediately after the laser was re-aligned to the second anapole, the EV was transported to the laser illumination and was trapped again. The second transport took ~5.8 seconds. Finally, the laser was turned off and the EV was released. (MOVIE)

**Supporting video 3 to 5: Trapping polystyrene beads using HAT system for various CTAC concentration.**

Anapole trapping of a single 20 nm polystyrene bead in DI water (SIV3), 0.5 mM (SIV4) and 1 mM (SIV5) CTAC solution. (MOVIE)

AUTHOR INFORMATION

**Corresponding Author**

*Justus C. Ndukaife: justus.ndukaife@vanderbilt.edu

**Author Contributions**

J.C.N. conceived and guided the project. C.H. fabricated the samples and performed the trapping experiments. I.H. measured the scattering spectra using a home-made dark-field measurement setup. C.H. and Y.J. conduct the simulations. J.C.N and C.H. discussed the results and wrote the manuscript.

ACKNOWLEDGMENT

The authors acknowledge financial support from the National Science Foundation NSF CAREER Award (NSF ECCS 2143836). The authors also express their gratitude to Vanderbilt Institution of Nanoscale Science and Engineering (VINSE) for their fabrication facilities.